\documentstyle [12pt] {article}
\begin{document}
\begin{titlepage}
\begin{center}
\vspace{2cm}
\LARGE
Hierarchical Clustering and the Butcher-Oemler Effect \\
\vspace{1cm}
\large
Guinevere Kauffmann \\
\vspace{0.5cm}
{\em Dept. of Astronomy, University of California, Berkeley, CA, 94720}\\
\vspace{0.8cm}
\end{center}
\begin {abstract}
We show that the rapid evolution in the fraction of blue, star-forming galaxies
seen in
clusters as a function of redshift (the Butcher-Oemler effect) can be explained
very
simply if structure formation in the universe proceeeds hierarchically. We show
that
a rich cluster observed at high redshift has had a significantly different
evolutionary
history to a cluster of the same richness observed today. High redshift
clusters
take longer to assemble and thus undergo more merging at small lookback times.
We have investigated
two models of star formation in cluster galaxies: 1) a model in which star
formation
is induced by galaxy-galaxy mergers and interactions and 2) a model in which
star formation is
regulated by the infall of galaxies onto larger systems such as groups and
clusters.
Both models produce trends consistent with the Butcher-Oemler effect.
Our models of cluster formation and evolution allow us to make predictions
about
trends in the observed properties of clusters with redshift. We find that there
should
be a correlation between the mass of the cluster or group and the strength of
the observed
Butcher-Oemler effect, with more massive systems exhibiting more evolution than
less massive
systems.
We also predict that both the blue galaxy fraction and the incidence of
interacting
or merging galaxies in rich clusters should rise continuously with redshift.
Finally, we have explored the influence of cosmological parameters on our
predictions
for cluster evolution. We find that models in which structure formation occurs
at
very early epochs, such as low $\Omega$ models, predict rather little recent
star formation
and merging activity in clusters at redshifts around 0.4.
Larger observational samples and more detailed modelling will be needed
before any firm constraints can be placed on cosmology.
\end {abstract}
\end {titlepage}
\normalsize

\section {Introduction}
It is now well established that substantial differences exist between nearby
clusters
and clusters observed at redshifts greater than 0.2. The increase in the number
of blue
galaxies that occur in concentrated galaxy clusters at high redshift is
commonly referred
to as the Butcher-Oemler effect, in honour of the two astronomers who first
presented
photometric evidence for this phenomenon (Butcher \& Oemler 1978).
Substantial spectroscopic follow-up work on the Butcher-Oemler clusters
has since been carried out (Dressler \& Gunn 1982,1983; Lavery \& Henry 1986;
Couch \& Sharples 1987, Charlot \& Silk 1994). These studies
have confirmed that many of the blue, star-forming galaxies are indeed cluster
members and not merely interlopers along the line-of-sight. In addition,
analyses
of the spectra of these galaxies have yielded important clues about the nature
of these objects. Approximately 50-60 \% of the population show strong emission
lines
of [OII], [OIII] and H$\beta$, indicative of vigorous star formation. The
remainder
show very strong Balmer absorption lines, but negligible emission lines. These
objects
also tend to be considerably redder.
It has been shown (Dressler \& Gunn 1983, Couch \& Sharples 1987) that many of
the
features of these spectra may be understood if these galaxies were subject to a
substantial burst of star formation at some point in their histories.
The very blue colours of some of the strong emission-line objects require star
formation
rates considerably higher than those observed in present-day spiral galaxies.
The
strong absorption-line galaxies have been postulated to be "post-starburst"
systems.
A post-starburst population would include a large population of A stars
contributing
to the strong Balmer lines, but if the star formation rate declined rapidly
following
the burst, it would lack the younger O and B stars that excite the
line-emitting gas
in HII regions.

The discovery of a sizeable population of star-forming galaxies in high
redshift clusters
has led to a great deal of speculation about the physical processes that could
be
responsible for this activity. One popular hypothesis is that mergers or close
encounters
between the cluster galaxies may be responsible. Lavery \& Henry (1988) and
Lavery, Pierce
\& McClure (1992) have presented evidence from ground-based imaging of distant
clusters
that a substantial fraction of the blue galaxies are either multiple systems or
have
peculiar morphologies and features, such as tidal tails, suggestive of
interactions or
mergers. Further support for the merging hypothesis has also come from the
first Hubble
Space Telescope (HST) observations of distant clusters, which suggest that
interactions
are responsible for many of the bluest galaxies with the most vigorous star
formation
(Dressler et al 1994, Couch et al 1994).
However, these observations also indicate that approximately half the blue
population is
composed of fairly ordinary-looking late-type spirals. Dressler \& Gunn (1983)
have
suggested that the infall of gas-rich galaxies into the high-pressure, dense
intracluster
medium might help trigger starbursts. These galaxies would then be observed as
luminous,
blue spiral systems. Although recent observational data has shed considerable
light
on the physical processes responsible for triggering star formation in cluster
galaxies,
no convincing explanation has yet been given as to {\em why} there should be
a trend in the properties of rich clusters with redshift. As put by Dressler
(1993),
``the reason(s) for the demise of these actively star-forming systems in the
cluster
environment by the present epoch remains the major mystery in this field''.

In this paper, we study cluster evolution within the framework of the
hierarchical
clustering picture of structure formation in the universe. In this picture
structure forms from
the bottom up, via the merging of small objects to form successively larger and
larger mass
systems. Considerable support for this theory has been accummulating in recent
years as
astronomical observations of galaxies and clusters have reached to higher
redshifts.
Perhaps the strongest support for
a bottom-up scenario of structure formation has come from X-ray surveys, which
find strong
evidence for a decline in the abundance of X-ray luminous clusters out to
redshifts of
$\sim 1$ (Edge et al 1990, Gioia et al 1990, Henry et al 1992, Castander et al
1993).
In addition, many clusters observed in the X-ray show evidence of substructure
(see
for example Jones \& Foreman 1992), indicating that they may have formed
recently by
the merging of a number of smaller subunits.
In this paper, we show that the trend of
more star-formation activity in rich clusters at high redshift may be explained
naturally within the framework of the hierarchical clustering picture.
We show that a cluster that is assembled at high redshift has had a
very different merging history to a cluster of the same mass assembled at the
present day.
High redshift
clusters take much longer to form. The increased star-formation
activity seen in the cluster galaxies can be explained very simply by the
increased merging activity taking place
in the dark matter component at an epoch just prior to the time the cluster is
observed
as single, collapsed object. We explore two simple models for star formation in
cluster galaxies. These models are  designed to reflect the
interaction-driven and infall-driven pictures
of star formation discussed above. We show that {\em both} models lead to
trends that
are consistent with the observations. We also explore the effect of our choice
of cosmological
initial conditions on our predictions for cluster evolution. This aim of this
paper is
to explain the {\em qualitative} trends seen in the data. A more detailed
quantitative
analysis is reserved for a later publication.

\section {The Merging History of Clusters}
We use the algorithm developed by Kauffmann \& White (1993, hereafter KW)
to trace the merging history of
a dark matter halo of specified mass present a redshift z. This algorithm is
based on
an extension of the Press-Schechter theory due to Bower (1991) and Bond et al
(1991).
In these papers, expressions were derived for the conditional probability that
material
in an object of mass $M_{1}$ at redshift $z_{1}$ would end up in an object of
mass $M_{0} > M_{1}$
at a later redshift $z_{0}$. The algorithm of KW used these probabilities to
construct
Monte-Carlo realizations of the merging histories of halos. These histories
were stored
in the form of a {\em tree} (akin to a family tree), with each successive layer
representing a step in redshift and branches indicating the merging paths
of the halo's progenitors. The reader is referred to KW for further details.

Let us now consider the merging histories of a set of clusters of fixed mass,
which
are observed at a series of redshifts: $z_{obs}=0, 0.07, 0.15, 0.2, 0.3$ and
$0.4$.
To do this, we trace the evolution of the mass of the {\em largest cluster
progenitor}
as a function of lookback time, t, from the the redshift $z_{obs}$. In figure
1(a)
we show results for a dark matter halo of mass $10^{15} M_{\odot}$, comparable
to
the mass of a moderately rich cluster. We adopt a cold dark matter (CDM)
cosmology
with $\Omega=1$, $H_{0}= 50$ km s$^{-1}$ Mpc$^{-1}$ and
bias parameter b=2.5 as our reference model. In section 4, we will discuss
the influence of other choices of cosmological initial conditions on our
predictions.

The curves in figure 1(a) represent averages over 50 Monte Carlo realizations
of the formation
history of a $10^{15} M_{\odot}$ cluster. As can be seen, the evolutionary
history is strongly
dependent on $z_{obs}$. There are two reasons for this effect. The first is
simply that
the universe is younger as measured in gigayears at higher redshift. The second
is that a rich
cluster observed at high redshift results from a much  higher amplitude
fluctuation in the initial
linear density field than a cluster of the same mass seen today. As a result,
high redshift
assemble considerably later and accrete
a much larger percentage of their final mass in the few gigayears just prior to
$z_{obs}$.
In figures 1(b) and 1(c), we show similar curves for halos with mass $10^{13}
M_{\odot}$ and $10^{12} M_{\odot}$ respectively, corresponding to the mass of a
galaxy
group and to that of a typical bright galaxy. It is clear that halos of lower
mass
form earlier, a trend which has  also been noted by Lacey \& Cole (1993). In
addition, we
also see the difference in the evolutionary history of high and low redshift
systems
is much smaller for halos of lower mass.

These results may be understood intuitively as follows. In a hierarchical
universe, the
characteristic mass scale of collapsed objects increases with time. Low mass
objects
typically form at higher redshift than high mass objects. In addition, the most
massive
halos present at any given time tend to form via the merging of smaller objects
that are
close to the characteristic mass at that epoch. This is why rich clusters,
which are on the high-mass
tail of the distribution of collapsed objects in the universe today, tend to
form rather late
and via the merging of many smaller, more typical halos. A cluster of the same
mass present
at high redshift is even further out on the tail of the mass distribution and
the effect is correspondingly more severe. Halos with masses corresponding to
those of galaxy groups
and isolated bright galaxies are much more typical objects at redshifts less
than 1,
and thus do not display these dramatic trends in their evolutionary history.
It is only at very high redshifts, when groups and galaxies become rare
objects, that
these effects become important.

We have seen that a trend in the evolutionary history of a rich cluster towards
later
assembly times with increasing $z_{obs}$,
is a {\em natural consequence} of hierarchical structure formation. It would
also be
natural to expect that the observed properties of the galaxies in high redshift
clusters should
reflect this trend in the dark matter evolution. Late assembly means that
considerable
merging of the dark matter component has occurred at late evolutionary times,
and it is likely that this merging activity would lead to
an increased level of star formation (ie the Butcher-Oemler effect) in the
cluster galaxies.
We will discuss this in more detail in the next section.
Another consequence of hierarchical clustering is that the strength of the
Butcher-Oemler
effect should correlate with the mass of the system. This is evident from
figures 1(b) and
1(c), which show that the difference in the evolutionary histories of low mass
halos
as a function of $z_{obs}$ is much smaller than that for massive halos. This
accords well
with recent observational studies of the evolution of galaxies in high redshift
groups selected
around radio galaxies (Allington-Smith et al 1993). These studies have shown
that groups
exhibit much weaker evolutionary trends with redshift than do rich clusters.

One very useful feature of our Monte-Carlo approach to calculating the merging
history of
dark matter halos is that we are able to obtain a measure of the scatter in the
evolutionary
history of individual clusters. Recall that the curves in figure 1 represent
averages over
50 Monte-Carlo realizations. In figure 2, we show individual merging histories
for 30
different clusters with mass $10^{15} M_{\odot}$ and $z_{obs}=0.4$. The thick
dashed line
shows the average merging history for a $10^{15} M_{\odot}$ cluster observed at
$z_{obs}=0$.
As can be seen, the scatter in individual merging histories is rather large and
it is
interesting to note that a small percentage of clusters with $z_{obs}=0.4$ have
evolutionary histories that match the $z_{obs}=0$ curve out to a lookback times
of
3-4 gigayears. If star formation in cluster galaxies is indeed fueled by recent
merging in the dark matter component of the cluster, we would expect that a
small fraction
of clusters present at $z=0.4$ to appear very similar to clusters seen today.
One well-known observational example is the z=0.5 cluster CL0016+16 which
contains few,
if any, blue
cluster members (Koo 1981). Larger observational samples will be needed before
any quantitative
comparison can be made with theory.

\section {Models for Star Formation in Cluster Galaxies}
In this section, we explore two simple models for the rate at which stars form
in cluster
galaxies. This first model is motivated by the ``infall'' picture of Dressler
\& Gunn (1983).
We assume that stars form at a constant rate in galaxies until they are
incorporated into
larger systems, such as groups or clusters. Once the galaxy has been accreted,
we assume that star
formation ceases because gas is stripped from the galaxy due to the
ram-pressure
of the intracluster medium. In this model, the epoch at which galaxies are
accreted by larger
systems delineates the transition between active star formation and the
``post-starburst''
phase discussed by Dressler \& Gunn (1983). For illustrative purposes, we will
assume
that a galaxy can form stars so long as it is contained in a halo with mass in
the range
$10^{11}-10^{13} M_{\odot}$. If figure 3, we show the mass fraction of the
cluster which
is contained in halos in this mass range as a function of lookback time for a
$10^{15} M_{\odot}$
cluster observed at $z_{obs}=0, 0.07, 0.15, 0.2, 0.3$ and $0.4$. It is clear
that
star formation occurs at earlier lookback times for clusters seen today than
for clusters
observed at high redshift. Again, this simply follows from the fact that high
redshift
clusters assemble later. We conclude that infall-driven star formation will
lead to a trend
consistent with the Butcher-Oemler effect.

The second model is motivated by the interactions/mergers picture favoured by
Lavery \&
Henry (1986) and Lacey \& Silk (1991) in which encounters between
close pairs of galaxies are responsible for the observed
star formation in cluster galaxies at high redshift. We assume that a merger
between
galaxy-sized halos of roughly equal mass results in a burst of star formation.
In figure
4, we show the rate at which galaxy-sized progenitors of a $10^{15} M_{\odot}$
cluster merge with each other as a function of lookback time, again for
clusters with
$z_{obs}= 0, 0.07,
0.15, 0.3, 0.3$ and $0.4$. We plot the fraction of the mass of the cluster
per gigayear in halos that undergo a merger with another object of at least
half
its mass, and we restrict the halos to be in the mass
range $10^{11}-10^{13} M_{\odot}$. Once again, we see that the inferred star
formation
rate peaks at smaller lookback times for clusters observed at high redshift
than
for clusters seen today. A higher incidence of interacting galaxies in high
redshift
clusters is thus also a natural outcome of hierarchical clustering theory. It
should be
noted that in our picture, merging occurs in lower mass groups {\em before} the
cluster is
assembled. Not much merging is believed to occur once the cluster has formed
because of
the high relative velocities between pairs of galaxies.

One popular idea is that mergers between galaxies of nearly equal mass lead to
the transformation
of disk galaxies into ellipticals or the spheroidal bulges
of spirals. We can transform the instantaneous merging rates shown in figure 4
into a
plot of the cumulative mass fraction of the cluster contained in merged
remnants
as a function of lookback time (figure 5). If we assume that a large fraction
of the stars
in an elliptical galaxy form in a burst during the merging event, we infer from
figure
5 that elliptical galaxies in high redshift clusters have substantially younger
stellar populations than cluster ellipticals seen today. This trend also seems
to be apparent
in the observational data. Aragon-Salamanca et al (1993) present evidence for a
systematic
bluing of the reddest galaxies in rich clusters as a function of redshift. This
trend
is in agreement with studies of the evolution of the amplitude of the 4000
\space \AA
break
in red cluster galaxies by Dressler \& Gunn (1990).

We have shown that both the infall-driven and merging-driven pictures of star
formation
lead to trends consistent with observations. In practice, both effects are
likely to be
important in explaining cluster evolution. We can address this question
explicitly by
comparing the mass fraction of the cluster tranformed into merger remnants
during the
final 3 gigayears of its evolutionary history,
with the mass fraction in galaxy-sized halos which have never merged, but have
been accreted
onto a larger system during the same time. We find that the two mass fractions
are
approximately equal for all values of $z_{obs}$. We conclude that {\em both}
the infall and
merging mechanisms are likely to be responsible for the
blue, actively star-forming galaxies seen
in rich clusters at all redshifts.

\section {Dependence on Cosmology}
As we have discussed, the trend in the evolutionary behaviour of rich clusters
with redshift
is intrinsic
to hierarchical clustering scenarios and occurs because rich clusters are on
the high mass
tail of the distribution of collapsed objects in the universe today. One of the
important
tests of any model of structure formation is that it be able to reproduce the
observed
abundance of rich clusters. So it should come as no surprise that the
trends we have discussed in the previous sections are much the same for most
realistic
choices
of cosmological parameters. What does turn out to be sensitive to the choice of
cosmological
model is the epoch at which processes such as galaxy merging and infall can be
expected
to occur.

Let us take, for example, the case of a low density ($\Omega=0.2$) CDM model.
Structure
forms considerably earlier in a low density universe. By the present day, the
growth of
perturbations has slowed considerably and much less merging
of dark matter halos will be taking place. In addition, the overall age of the
universe
is a factor of a third larger. In figure 6, we show the cumulative mass
fraction of
a $10^{15} M_{\odot}$ cluster in merger remnants as a function of lookback
time.
The curves in this figure are calculated exactly the same way as described for
figure 5.
Comparing figure 6 to figure 5, we see that all the curves are now shifted to
much
higher lookback times. In an open universe elliptical galaxies
form considerably earlier and there is much less merging going on at late
evolutionary times
in the cluster. At an observed redshift of 0.4, only 2\% of the
cluster mass has been involved in a major merging event
over the past 3 gigayears in the $\Omega=0.2$ CDM model, as opposed to 16\% for
the
$\Omega=1$ model. In figure 7, we show the prediction for the fraction of the
cluster in
star-forming units in the ``infall'' picture discussed in section 3. This
diagram is the
analogue of figure 3. We see that star formation in the cluster now peaks at
much higher
lookback times and that even at $z_{obs}=0.4$, relatively little star formation
occurs
during the last few gigayears of the cluster's history.
Conversely, models with late structure formation, such as mixed dark
matter models, lead to a prediction of more recent merging and star formation
activity
in cluster galaxies. One obvious way to constrain cosmological parameters,
is to make use of the new high resolution imaging techniques to
determine, as a function of redshift,
what fraction of bright cluster galaxies can be classified as
interacting systems, and compare this with our predicted halo merging
rates. Observational studies along these lines are currently underway. Another
sensitive
test of cosmology would be the observed colour evolution of elliptical galaxies
in
clusters, especially at high redshift. Models with late structure formation
such
as MDM or high-bias CDM are predict that at redshifts greater than 1, rich
clusters
should contain almost no luminous elliptical galaxies with old stellar
populations.

\section {Conclusions}
We have shown that the rapid evolution in the fraction of blue, star-forming
galaxies seen in rich
clusters between the present day and redshifts of around 0.4
can be explained very simply if structure
formation in the universe occurs hierarchically. In hierarchical clustering
models,
structure evolves to form larger and larger mass systems with time. A rich
cluster observed at $z=0.4$ {\em cannot} be regarded as the direct progenitor
of a cluster of
similar richness seen today. High redshift rich clusters are much rarer systems
and
consequently have had different evolutionary histories to those of rich
clusters today.
They assemble later and undergo more merging at late times. We have explored
the predictions
of two models of star formation in cluster galaxies:
\begin {enumerate}
\item A model in which star formation is triggered by galaxy interactions and
mergers.
\item  A model in which star formation occurs only in isolated galaxies and
ceases once
galaxies are accreted onto larger systems such as groups and clusters.
\end {enumerate}
We show that both models lead to trends consistent with the observed
Butcher-Oemler effect.
We also show that in the hierarchical picture, both galaxy infall and merging
are important
during the last few gigayears of cluster evolution. The observed star formation
in high
redshift cluster galaxies is most likely explained by both mechanisms.
We have also shown that it is a prediction of hierarchical clustering theory
that the strength
of the Butcher-Oemler effect should correlate with the mass (or richness) of
the galaxy
cluster or group, with more massive systems exhibiting a larger effect. In
addition,
the strength of the Butcher-Oemler effect should increase with redshift.
Finally, we have explored the effect of the choice of cosmological initial
conditions on
our predictions for cluster evolution. The most dramatic differences are found
in low-density
models, which predict considerable less recent star formation and merging
activity in
rich clusters at {\em all} redshifts, than models with $\Omega=1$.

There is no doubt that observational data on high redshift galaxy clusters will
improve
very significantly over the next few years. High resolution imaging of clusters
by the
recently repaired Hubble Space Telescope is currently underway.
These studies will provide much more reliable statistical information about
cluster galaxy properties, such as morphology, and the frequency of the
incidence of
interacting systems in clusters. On the theoretical side, the next obvious step
would be to combine
our Monte-Carlo models of cluster formation with spectrophotometric population
synthesis
models in order to make detailed comparisons with the data. It seems promising
that
this combination of approaches will help constrain cosmological models and
significantly
improve our understanding of the physical processes that regulate galaxy
formation in
clusters.
\\

\vspace{0.8cm}

\large
{\bf Acknowledgments}\
\normalsize
I thank Joe Silk, Stephane Charlot, Richard Ellis and Simon White for helpful
discussions.

\pagebreak
\Large
\begin {center} {\bf References} \\
\end {center}
\vspace {1.5cm}
\normalsize
\parindent -7mm
\parskip 3mm

Allington-Smith, J.R., Ellis, R.S., Zirbel, E.L. \& Oemler, A., 1993, APJ, 404,
521

Aragon-Salamanca, A., Ellis, R.S., Couch, W.J. \& Carter, D., 1993, MNRAS, 262,
764

Bond, J.R., Cole, S., Efstathiou, G. \& Kaiser, N., 1991, APJ, 379, 440

Bower, R., 1991, MNRAS, 248, 332

Butcher, H.R. \& Oemler, A., 1978, APJ, 219, 18

Castander, F.J., Ellis, R.S., Frenk, C.S., Dressler, A. \& Gunn, J.E., 1994,
preprint

Charlot, s. \& Silk, S., 1994, ApJ, in press

Couch, W.J. \& Sharples, R.M., 1987, MNRAS, 229, 423

Couch, W.J., Ellis, R.S., Sharples, R.M. \& Smail, I, 1994, APJ, in press

Dressler A. \& Gunn, J.E., 1982, APJ, 263, 533

Dressler, A. \& Gunn, J.E., 1983, APJ, 270, 7

Dressler, A. \& Gunn, J.E., 1990, in Kron, R.G. ed, ASP Conference Series Vol.
10,
Evolution of the Universe of Galaxies: Edwin Hubble Centennial Symposium, ASP,
San Francisco, p208

Dressler, A., 1993, in Chincarini, G., Iovino, A., Maccacaro,T. \& Maccagni,
D., eds,
Observational Cosmology, San Francisco:ASP, p225

Dressler, A., Oemler, A., Butcher, H.R. \& Gunn, J.E., 1994, APJ, in press

Edge, A.C., Stewart, G.C., Fabian, A.C. \& Arnaud, K.A., 1990, MNRAS, 245, 559

Gioa, I.M., Henry, J.P., Maccacaro, T., Morris, S.L., Stocke, J.T. \& Wolter,
A.,
1990, APJ, 356, L35

Henry, J.P, Gioa, I.M., Maccacaro, T., Morris, S.L., Stocke, J.Y. \& Wolter,
A.,
1992, APJ, 386, 408

Jones, C. \& Foreman, W., 1992, in Fabian, A.C., ed, Clusters and Superclusters
of Galaxies,
Kluwer, p49

Kauffmann, G. \& White, S.D.M., 1993, MNRAS, 261, 921

Koo, D.C., 1981, APJ, 251, L75

Lacey, C.G. \& Silk, J., 1991, ApJ, 381, 14

Lacey, C.G. \& Cole, S., 1993, MNRAS, 262, 627

Lavery, R.J. \& Henry, J.P., 1986, APJ, 304, L5

Lavery, R.J., Pierce, M.J., McClure, R.D., 1992, AJ, 104, 1067

\pagebreak

\Large
\begin {center} {\bf Figure Captions} \\
\end {center}
\vspace {1.5cm}
\normalsize
\parindent 7mm
\parskip 8mm

{\bf Figure 1:} The evolution of the mass of the most massive progenitor of a
halo as a
function of lookback time. We plot the logarithm of the mass of the progenitor
divided
by the mass of the halo as a function of lookback time in gigayears, for halos
of fixed mass that are
assembled at redshifts 0, 0.07, 0.15, 0.2 0.3 and 0.4.
The curves represent averages over 50 Monte-Carlo realizations of the halo
merging
history. We show results for halos of three different masses.
a) $10^{15} M_{\odot}$ b)$10^{13} M_{\odot}$ c)$10^{12} M_{\odot}$ .

{\bf Figure 2:} The scatter in the evolutionary history of a $10^{15}
M_{\odot}$
cluster. As in figure 1, we show the evolution of the mass
of the largest cluster progenitor. The thin lines show 30 individual
evolutionary histories for
clusters assembled at z=0.4.
The thick solid line shows the {\em average} evolutionary history of a cluster
assembled at z=0.4.
The thick dashed line shows the average evolutionary history of a cluster with
the same mass that is assembled today.

{\bf Figure 3:} The mass fraction of a $10^{15} M_{\odot}$ cluster that is in
halos
with masses in the range $10^{11}-10^{13} M_{\odot}$, plotted  as a function of
lookback time.
Solid, dotted, short-dashed, long-dashed, short-dashed-dotted and
long-dashed-dotted
curves are for clusters observed at redshifts 0, 0.07, 0.15, 0.2, 0.3 and 0.4,
respectively.

{\bf Figure 4:} The mass fraction (per gigayear) of a $10^{15} M_{\odot}$
cluster
contributed by halos undergoing a merger with objects of at least half their
mass, plotted as a function of lookback time.
Halos have masses in the range $10^{11}-
10^{13} M_{\odot}$. Explanation of the line types is as given in figure 3.

{\bf Figure 5:} The cumulative mass fraction of a $10^{15} M_{\odot}$ cluster
contributed by halos in the mass range $10^{11}-10^{13} M_{\odot}$ that have
undergone major mergers, plotted as a function of
lookback time. Explanation of the line types is as given in figure 3.

{\bf Figure 6:} The cumulative mass fraction of a cluster in merger
remnants (as in figure 5), plotted as a function of lookback time
for a cold dark matter model with $\Omega=0.2$.
Explanation of the line types is as given in figure 3.

{\bf Figure 7:} The mass fraction of a cluster in galaxy-sized halos (as in
figure 3), plotted
as a function of lookback time for a cold dark matter model with $\Omega=0.2$.
Explanation of the line types is as given in figure 3.
\pagebreak

\end {document}